\documentclass[twocolumn,prb]{revtex4-2}
\usepackage{titlesec}
\usepackage{subfigure}
\usepackage{verbatim}
\usepackage{amsmath}
\usepackage{amsfonts}
\usepackage{amssymb}
\usepackage{feynmp}
\usepackage{graphicx}
\usepackage{setspace}
\usepackage{comment}
\usepackage{dsfont}
\usepackage{color}
\usepackage{bm}
\usepackage[pdftex,colorlinks=true,linkcolor=blue,citecolor=blue]{hyperref}
\usepackage{float}
\usepackage{enumerate} 
\begin{document}
\title{Topological band insulators without translational symmetry}
\author{Shuo Wang}
\author{Jing-Run Lin}
\author{Zheng-Wei Zuo}
\email{zuozw@haust.edu.cn}
\affiliation{School of Physics and Engineering, Henan University of Science and Technology, Luoyang 471023, China}
\date{\today}
\begin{abstract}
In the research of the topological band phases, the conventional wisdom is to start from the crystalline translational symmetry systems. Nevertheless, the translational symmetry is not always a necessary condition for the energy bands. Here we propose a systematic method of constructing the topological band insulators without translational symmetry in the amorphous systems. By way of the isospectral reduction approach from spectral graph theory, we reduce the structural-disordered systems formed by different multi-atomic cells into the isospectral effective periodic systems with the energy-dependent hoppings and potentials. We identify the topological band insulating phases with extended bulk states and topological in-gap edge states by the topological invariants of the reduced systems, density of states, and the commutation of the transfer matrix. In addition, when the building blocks of the two multi-atomic cells have different number of the lattice sites, our numerical calculations demonstrate that the existences of the flat band and the macroscopic bound states in the continuum in the amorphous systems. Our findings uncover an arena for the exploration of the topological band states beyond translational symmetry systems paradigm.
\end{abstract}

\maketitle

\emph{Introduction}---
Over the past two decades, the study of symmetry-protected topological quantum materials  has gained significant attention in view of their exotic electronic phenomena on the fields of condensed matter\cite{Hasan10RMP, QiXL11RMP, Bernevig13Book, Franz15RMP, ChiuCK16RMP, AsbothJ16, BansilA16RMP, WenXG17RMP, BergholtzEJ21RMP, MoessnerR21Book}. The exploration of topological matter has been rigorously extended to diverse artificial fields such as photonic systems\cite{Ozawa19RMP}, and phonon systems\cite{ZhuWW23RRP}. Much of understanding of the topological quantum states is based on crystalline band theory because of the translational symmetry. According to the crystallographic group theory, the topological band states in crystalline solids could be identified by the symmetry-indicator theory\cite{PoHC17NTC, KruthoffJ17PRX,  WatanabeH18SA} and topological quantum chemistry\cite{BradlynB17NT,ElcoroL21NTC}.

As we know, these topological band states are stable against weak disorder. However, when the translational symmetry breaks down, the aperiodic systems such as disordered and amorphous systems have their own unique topological states distinct from the topological crystalline systems. For the disordered systems, the strong disorder could lead the system from topologically trivial states to topological Anderson states\cite{LiJ09PRL, GrothCW09PRL, GuoHM10PRL, XingYX11PRB, MondragonShem14PRL, AltlandA14PRL2, MeierEJ18SCI, StuetzerS18NT, LiuGG20PRL, CuiXH22PRL, ZuoZW22PRA, ChenR23PRB, ZhangH23PRB, LoioH24PRB,PadhanvA24PRB, ZuoZW24PRB}.  Based on the self-consistent Born approximation, the topological Anderson phase transitions could be understood by the effective medium theory \cite{GrothCW09PRL} with  renormalization of system parameters, including the mass term, hopping term, and potential, etc. On the other hand, structural disorder could  derive the amorphous systems from the topologically trivial states to various topological amorphous states\cite{AgarwalaA17PRL, ManshaS17PRB, MitchellNP18NTP, PoyhonenK18NTC, YangYB19PRL, YangB19PRB, CostaM19NanoLett, ChernGW19EPL,HuangHQ20PRB,  MarsalT20PNAS, MukatiP20PRB, SahlbergI20PRR, IvakiMN20PRR, ZhouPH20LSA, CorbaeP21PRB, WangJH21PRL, LiK21PRL,SpringH21SPP, WangCT22PRL, ChengXY23PRB, ZhangZ23SA, MarsalQ23PRB, Munoz-SegoviaD23PRR, PengT24PRB, RegisV24PRB, PengT25PRB} such as topological amorphous insulators, topological amorphous metals, and high-order  topological amorphous insulators.

The fundamental hypothesis to analyze the topological band phases is the translational symmetry. Now the following questions naturally arise: Is the translational symmetry a necessary condition for the topological band states? When the translational symmetry breaks down, could the topological band states emerge in the aperiodic systems? In this study, we try to answer these questions. In fact, translational symmetry is a sufficient but not necessary condition for the topological band states. In the following, we demonstrate the counterintuitive scenario that the topological band insulators emerge in the certain amorphous systems and propose a systematic and flexible approach to engineer the topological band insulators without the translational symmetry. We take the quasi one-dimensional (1D) amorphous systems as examples, where the two different building blocks of multi-atomic cells are arranged in a random fashion. These quasi-1D amorphous systems could be mapped onto the 1D clean generalized dimerized systems via the isospectral reduction (ISR) approach from spectral graph theory\cite{BunimovichL14Book, BarrettW17LAA, SmithD19PA, KemptonM20LAA}. Under the specific parameters, we illustrate the emergence of the topological band phases and topological edge states by means of the topological invariants, density of states, and transfer matrix method. This opens up the perspective of design the topological band insulators in the amorphous systems.

\begin{figure}[tbp]
\centering 
\includegraphics[width=0.46\textwidth]{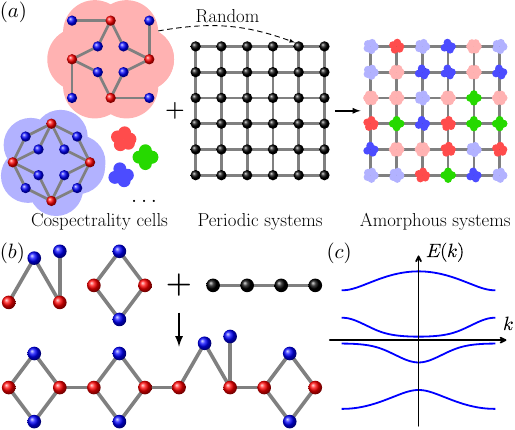}
\caption{$(a)$ The systematic construction of the topological band insulators without translational symmetry. $(b)$ Amorphous chain formed by tetratomic and diamond-shaped cells. $(c)$ The four topological Bloch bands of the amorphous chain.}
\label{FigConstruction}
\end{figure}

\begin{figure}[tbp]
\centering 
\includegraphics[width=0.46\textwidth]{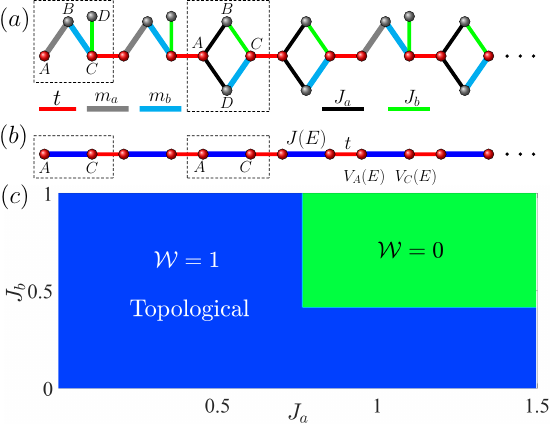}
\caption{The lattice structure and phase diagram of the diamond-tetratomic model. $(a)$ A typical configuration of the amorphous chain formed through the random arrangement of tetratomic and diamond cells. $(b)$ The reduced dimerized chain by ISR approach. 
$(c)$ Phase diagram of the original diamond-tetratomic amorphous system at $1/4$ filling in the $J_a-J_b$ parameter space.
}\label{FigDiamondPhase}
\end{figure}

\emph{Method}---
 Here, we provide the detailed framework for the construction of the topological band insulators in the amorphous systems. Firstly, the different multi-atomic cells (clusters) as the building blocks are utilized to introduce the structural disorder in the crystal lattice structures via amorphization methods such as the lattice sites substitutions and random bonding arrangements. For these amorphous systems, the conventional topological tools cannot directly characterize and classify the topological band features from the original system. Through the ISR approach, we divide the lattice sites in the amorphous systems into two classes: a structural set $S$ (retained sites) and its complement $\overline{S}$ (eliminated sites). Then, the isospectral effective reduced amorphous systems with the energy-dependent hoppings and potentials could be obtained. By tuning the system parameters appropriately, one could make sure that the reduced different cells become the cospectrality cells, where the matrix power for the block Hamiltonian of these reduced different cells $\left[H_i^n\right]_{SS}$ and $\left[H_j^n\right]_{SS}$ are identical for all non-negative integers $n$. From the quantum transport viewpoint, the cospectrality cells mean that the string transfer matrix of these reduced different cells are commutated. As a result, the original amorphous system becomes a periodic crystal system with energy-dependent hoppings and potentials. Then, according to the topological band theories such as symmetry-indicator theory and topological quantum chemistry, we can identify the topological band states and phase transitions in the original amorphous systems. Figure \ref{FigConstruction}$(a)$ illustrates the general workflow of the construction framework via lattice sites substitutions. The specific quasi-1D amorphous system [the detailed parameters of the couplings, see Fig.\ref{FigDiamondPhase}$[a]$] with four Bloch bands is given in Fig.\ref{FigConstruction}$(b)$ and Fig.\ref{FigConstruction}$(c)$.
 
\emph{The diamond-tetratomic model}---
As a warm up, we consider the quasi-1D system of noninteracting particles with structural disorder (see Fig.\ref{FigDiamondPhase}$[a]$), where the tetratomic and diamond-shaped building blocks are assigned at random with equal probability(the different probabilities lead to the same results). The tetratomic (diamond) cell is composed of four different lattice sites, which are marked as $A$, $B$, $C$, and $D$. The inter-cell coupling amplitude is a fixed constant $t$. In the tetratomic (diamond) cell, $m_a, m_b$, and $J_b$ ($J_a, J_b$, and $m_b$) are nearest-neighbor hopping amplitudes between the red lattice sites and gray lattice sites. For the general coupling parameters, the system would have the discrete energy spectrum with localized bulk states. Here, for simplicity, we assume that the parameters $m_a=\sqrt{2}J_a$ and $m_b=(\sqrt{2}+1)J_b$ (see the transfer matrix section for explanation). 

Next, we utilize the ISR approach to analyze this amorphous system (the ISR approach has proven applicable to diverse physical systems\cite{RontgenM20PRA, RontgenM21PRL, MorfoniosCV21PRB, RontgenM23PRL, ZhengLY23PRB, RontgenM24PRB, EekL24PRB,LinJR25CPB, EekL25PRL, MoustajA25PRB, LumenE25SPP, GuoQH25PAAP, LinJR25arXIv}).  All the lattice sites could be partitioned into two classes: a structural set $S$ (retained sites such as $A, C$) and its complement $\overline{S}$ (eliminated sites $B$ and $D$). The Hilbert space of Hamiltonian $H$ has the block matrix form: 
\begin{eqnarray}
H=\left[
\begin{matrix}
H_{SS} &  H_{\overline{S} S }   \\
H_{S \overline{S} } & H_{\overline{S}\overline{S}}  \\
\end{matrix}
\right],\label{Hcell}
\end{eqnarray}
where ${S}$ is the structural set, $\overline{S}$ is complement of ${S}$.  The effective reduced Hamiltonian $\widetilde{H}_S(E)$ over the structural set is given by:
\begin{equation}
	\widetilde{H}_S(E)=H_{SS}+ H_{S\overline{S}}(EI-H_{\overline{S} \overline{S}})^{-1}H_{\overline{S}S,}
		\label{ResH}
\end{equation}
where $I$ is the identity matrix.

As shown in Fig.\ref{FigDiamondPhase}$(a)$, we have colored the lattice sites differently to distinguish between structural set and complement set, which are marked as red circles ($A, C$) and gray circles ($B, D$), respectively.  According to the Eq.\ref{Hcell} and Eq.\ref{ResH}, we can express the isospectral effective Hamiltonian of the dimerized chain (Fig.\ref{FigDiamondPhase}$[b]$) derived by the diamond-tetratomic model as
\begin{eqnarray}
\widetilde{H}_S(E)=&\sum\limits_{j}^{N}J(E)c^\dagger _{A,j} c_{C,j}+\sum\limits_{j}^{N-1}t c^\dagger _{C,j} c_{A,j+1}+h.c.\notag\\
&+\sum\limits_{j}^{N}\left(V_{A}(E) c^\dagger _{A,j} c_{A,j}+ V_{C}(E) c^\dagger _{C,j} c_{C,j}\right), 
\label{eqirs}
\end{eqnarray}
where where $c_{\alpha, j}^{\dagger}$ ($c_{\alpha, j}$) denotes the creation (annihilation) operator at the site $\alpha$ ($\alpha$ stands for the lattice sites $A$ or $C$) of the $j$-th unit cell. The $N$ is the number of unit cell. Thus, the tetratomic and diamond cells have the same energy-dependent coupling amplitudes and on-site potentials, which are expressed as
\begin{eqnarray}
\left\{
\begin{aligned}
&J(E)=\frac{(2+\sqrt{2})J_{a}J_{b}}{E} \\
&V_{A}(E)=\frac{2J_{a}^{2}}{E}\\
&V_{C}(E)=\frac{(4+2\sqrt{2})J_{b}^{2}}{E}\\
\end{aligned}
\right. .
\label{energy-dependent-term}
\end{eqnarray}

Thus, the structural set ($A, C$ sites) of the tetratomic and diamond cells are cospectral sites (vertices) \cite{KemptonM20LAA, MorfoniosCV21LAA}, which means that the matrix power for the block Hamiltonian of the tetratomic and diamond cells $\left[H_{tet}^n\right]_{SS}$ and $\left[H_{dia}^n\right]_{SS}$ are identical for all non-negative integers $n$. According to the conditions of the cospectral sites, we could derive the relations $m_a=\sqrt{2}J_a$ and $m_b=(\sqrt{2}+1)J_b$. Now, we can see that this amorphous system reduces to a dimerized perfectly periodic lattice chain with the energy-dependent hoppings and potentials. After applying the Fourier transform under the periodic boundary condition (PBC), we can express the bulk Hamiltonian (set lattice constant of unit cell $a=1$) as
\begin{eqnarray}
\widetilde{H}_S(E, k)=&\left[
\begin{matrix}
V_{A}(E) &  J(E)+te^{-ik}   \\
J(E)+te^{ik} & V_{C}(E)  \\
\end{matrix}
\right]\notag\\ 
=&V(E)I+\left[
\begin{matrix}
u(E) &  J(E)+te^{-ik}   \\
J(E)+te^{ik} & -u(E) \\
\end{matrix}
\right],
\label{matrix2.1}
\end{eqnarray}
where the staggered onsite potentials $u(E)=\frac{1}{2}(V_{A}(E)- V_{C}(E))$, and $V(E) = \frac{1}{2}(V_{A}(E)+V_{C}(E))$. The analytic expressions of the energy dispersion is given by
\begin{eqnarray}
E^4-(2J_a^2+(4+2\sqrt{2})J_b^2+t^2)E^2-&\notag\\(4+2\sqrt{2})J_{a}J_{b}t\cos(k)E+2J_{a}^2J_{b}^2&=0. \label{matrix3}
\end{eqnarray}

Notably, $H_{1}(E, k)=\widetilde{H}_S(E, k)-V(E)I$ takes on the mathematical form of the Rice-Mele model\cite{RiceMele82PRL, AsbothJ16} with the energy-dependent hopping and potentials, which is the nontrivial and unexpected result. In momentum space, the isospectral effective Hamiltonian $H_{1}(k)$ is  formulated as $H_{1}(E, k) = \bm{h}(k)·\bm{\sigma} $, where $\bm{\sigma} = (\sigma_{x}, \sigma_{y} , \sigma_{z} )$ is the Pauli vector and $\bm{h}(k)$ = $(h_x , h_y, h_z)$ is composed of $h_x=J(E) + t\cos(k)$, $h_y=t\sin(k)$, $h_z=u(E)$. The topological features could be characterized by the winding number\cite{HattoriK23PRB}
\begin{equation}
\mathcal{W}=\frac{1}{2\pi}\int_{0}^{2\pi} \frac{ \partial \phi  }{ \partial k } \,dk
\label{eq2.3},
\end{equation}
where $\phi=\tan^{-1}({h_{y}}/{h_{x}} )$ is the azimuthal angle. As a result, we can investigate the topological properties of the original amorphous system under certain parameters range via the reduced clean Rice-Mele model. Figure \ref{FigDiamondPhase}$(c)$ shows the phase diagram  for the amorphous model with inter-cell coupling $t=1$ as a function of  the $J_a$ and $J_b$ parameters space at $1/4$ filling. The blue region ($\mathcal{W} =1$) in the phase diagram indicates the existence of the topologically nontrivial insulator phase. The green region ($\mathcal{W}=0$) represents the amorphous system is in trivial insulator phase.

\begin{figure}[tbp]
\centering 
\includegraphics[width=0.48\textwidth]{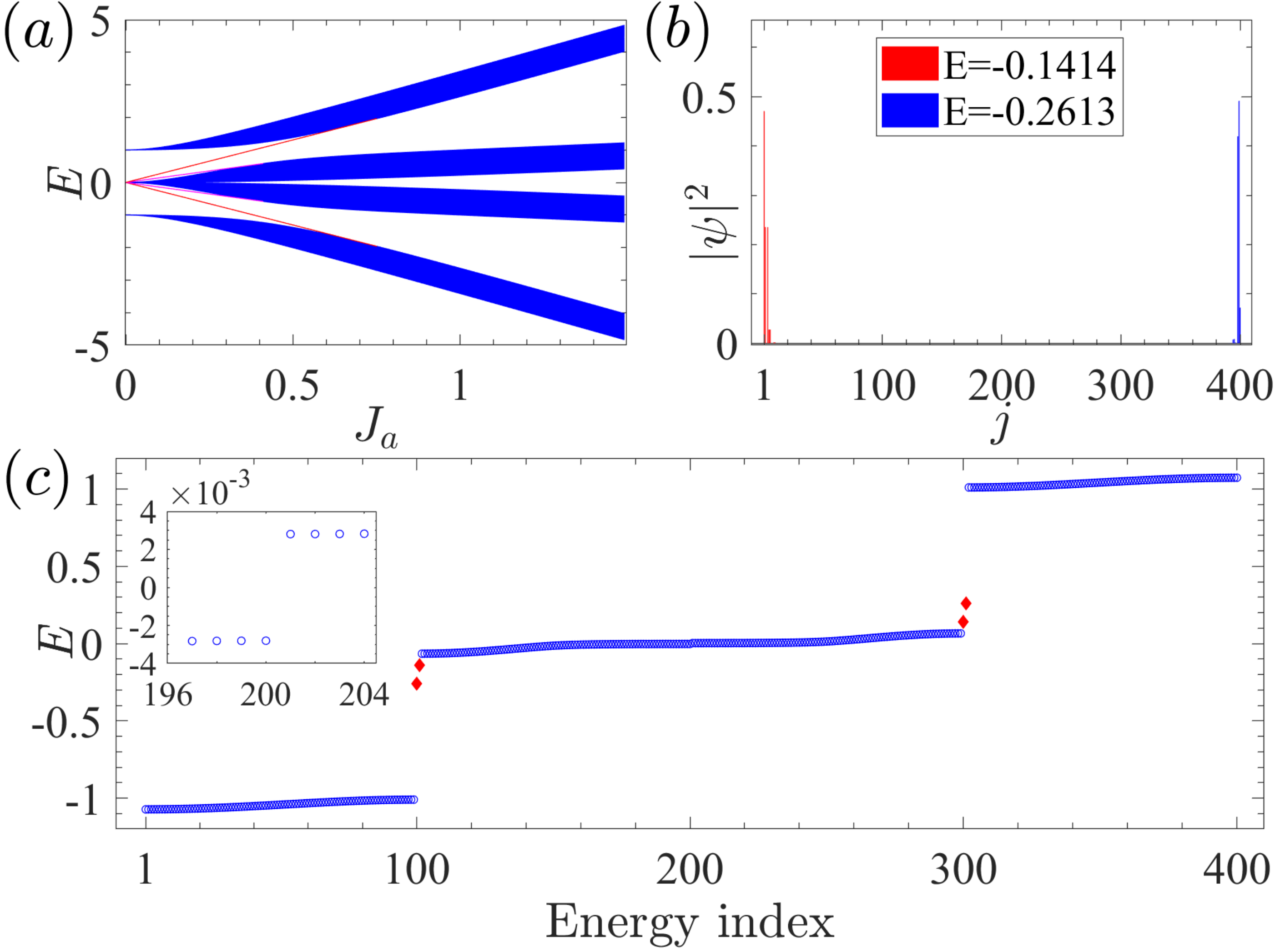}
\caption{ Real-space energy spectra and wave functions with the number of building cells $N=100$ and inter-cell hopping amplitude $t=1$. $(a)$ Energy spectrum as a function of intra-cell hopping amplitude $J_{a}$ ($J_{a}=J_{b}$) under OBC, $J_a$ from 0 to 1.5. $(b)$ The density profiles of the wave functions of two topological edge states at $1/4$ filling with energies $E_1=-0.1414$, $E_2 =-0.2613$ both marked as red circle in $(c)$. 
 $(c)$ Energy spectrum of the original diamond-tetratomic system for intra-cell hopping amplitudes $J_{a}=J_{b} =0.1$.}
\label{FigDiamondEdgeStates}
\end{figure}

The nontrivial topological invariant $\mathcal{W}$ is intimately related to topological edge states through the bulk-boundary correspondence. The energy spectrum of the finite system with open boundary condition(OBC) is presented in Fig.\ref{FigDiamondEdgeStates}$(a)$, for the parameters $J_a = J_b$, $t = 1$, and $N = 100$.  The red lines indicate the topological edge states of topological band insulators. In Fig.\ref{FigDiamondEdgeStates}$(c)$, the energy spectrum under OBC is plotted when the system parameters are $J_a = J_b=0.1$, $t = 1$, and $N = 100$. As shown in Fig.\ref{FigDiamondEdgeStates}$(b)$, the two topologically nontrivial in-gap edge states at $1/4$ filling are clearly localized in the left and right boundaries. It is worth mentioning that the topological edge states are robust against the weak disorder and the small deviations from these ratios $m_a = \sqrt{2}J_a, m_b=(1 +\sqrt{2})J_b$.

\begin{figure}[tbp]
\centering 
\includegraphics[width=0.48\textwidth]{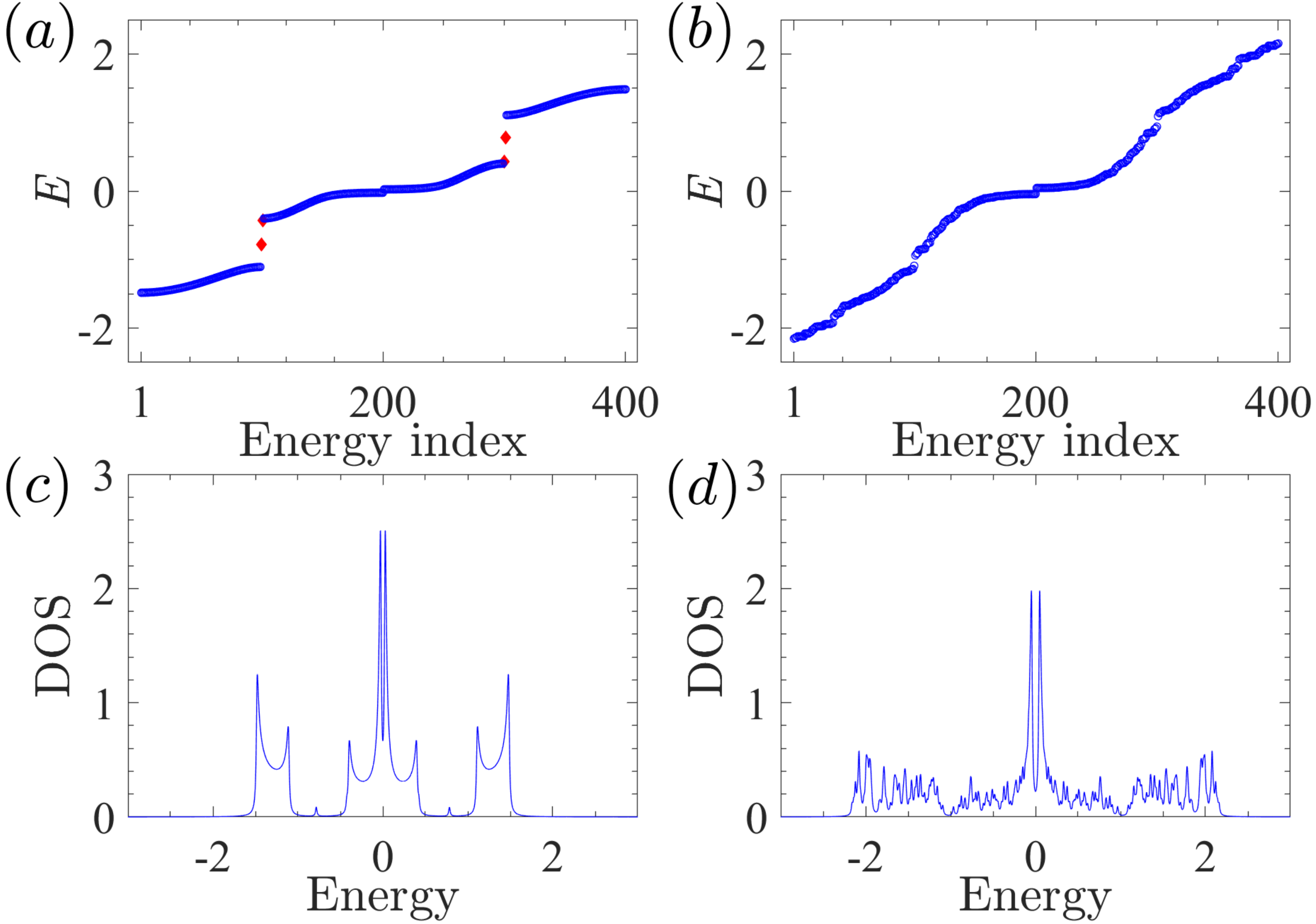}
\caption{ Energy spectrum and average DOS of the original diamond-tetratomic system under OBC with the system size $N = 100$, intra-cell couplings $J_a=J_b=0.3$, and the inter-cell hopping $t=1$. $(a)$ Energy spectrum with $m_a=\sqrt{2}J_a$ and $m_b=(\sqrt{2}+1)J_b$. $(b)$ Energy spectrum with $m_a=4J_a$ and $m_b=3.5J_b$ under one disorder configuration. Panels $(c)$ and $(d)$ are average DOS acting as a function of energy.
}
\label{FigEnergyDOS}
\end{figure}

\textit{DOS and transfer matrix method}---
In order to uncover the Bloch band characters, we use the average density of states (DOS) and transfer matrix method\cite{PalB13EPL, BiswasS23PRB} to verify the all extended bulk states and continuous energy spectrum. Fig.\ref{FigEnergyDOS} shows energy spectrum and average DOS for different ratios of the parameters describing the Hamiltonians of building blocks,  where system size $N=100$, intra-cell hopping amplitudes $J_a=J_b=0.3$, and the inter-cell coupling amplitude $t=1$. For the energy spectra of Fig.\ref{FigEnergyDOS}(a, b), the parameters intra-cell hopping $m_a=\sqrt{2}J_a$, $m_b=(\sqrt{2}+1)J_b$ and $m_a=4J_a$ and $m_b=3.5J_b$, respectively. The average DOS as a function of energy $E$ are shown in Fig.\ref{FigEnergyDOS}(c) and Fig.\ref{FigEnergyDOS}(d). It can be easily seen that the four parts of the energy spectrum in Fig.\ref{FigEnergyDOS}(a, c) are continuous and Fig.\ref{FigEnergyDOS}(b, d) show a discrete energy spectrum (the system is in non-topological phase region).

To analyze the extended properties of the bulk eigenstates, we assume that the intra-cell hoppings $m_a=\lambda_1 J_a$ and $m_b=\lambda_2 J_b$. Based on the Hamiltonian Eq.\ref{eqirs}, we can obtain the amplitudes of the eigenstates on neighboring lattice sites through a matrix equation
\begin{eqnarray}
\left(\begin{array}{c}
\psi_{n+1} \\
\psi_{n}  
\end{array}\right)
=M_n
\left(\begin{array}{c}
\psi_{n} \\
\psi_{n-1} 
\end{array} \right),
\label{transfer}
\end{eqnarray}
and the transfer matrix $M_n$ is given by 
\begin{eqnarray}
M_n= \left(\begin{array}{cc}
\frac{E-V_n(E)}{t_{n,n+1}} & \frac{-t_{n,n-1}}{t_{n,n+1}} \\ 
1 & 0 
\end{array} \right).
 \end{eqnarray}
 where $t_{n,n-1}$ ($t_{n,n+1}$) is the effective nearest-neighbor hopping between lattice sites $n$ and $n-1$ ($n$ and $n+1$) of the reduced  dimerized chain. The $V_n(E)$ is the on-site potential at lattice site $n$. For the diamond and tetratomic cells, there are four kinds of transfer matrices: $M_A$, $M_C$, $M_A'$, and $M_C'$ in the reduced dimerized chain. These transfer matrix elements are dependent on the on-site potentials and the nearest-neighbor hopping integrals of diamond and tetratomic cells, which are given by
 \begin{eqnarray}
&\text{Tetratomic cell }
&\left\{
\begin{aligned}
&J(E)=\frac{\lambda_1 \lambda_2 J_{a}J_{b}}{E} \\
&V_{A}(E)=\frac{(\lambda_1 J_{a})^{2}}{E}\\
&V_{C}(E)=\frac{(1+\lambda_2^2)J_{b}^{2}}{E}\\
\end{aligned}
\right.,
\end{eqnarray}
\begin{eqnarray}
&\text{Diamond cell }
&\left\{
\begin{aligned}
&J'(E)=\frac{(1+\lambda_2)J_{a}J_{b}}{E} \\
&V_{A}'(E)=\frac{2J_{a}^{2}}{E}\\
&V_{C}'(E)=\frac{(1+\lambda_2^2)J_{b}^{2}}{E}\\
\end{aligned}
 \right..
\end{eqnarray}

Now, we could define the unimodular transfer matrix of the tetratomic cell is $M_{tet}=M_AM_{C}$, and the unimodular transfer matrix of the diamond cell is $M_{dia}=M_A'M_{C}'$. Remarkably, with the parameters $\lambda_1=\sqrt{2},\lambda_2=1+\sqrt{2}$, we can obtain the commutator $[M_{tet},M_{dia}]=0$ for any energy $E$. Thus, for a typical realization of the amorphous chain, the commutation relation allows one to rearrange the string of the transfer matrix $\prod_j M_j$ as
\begin{equation}
\prod_j^\infty M_j=\left( \prod_j^\infty M_{j,tet} \right) \left( \prod_j^\infty M_{j,dia} \right).
\end{equation}

The right-hand side of the string of transfer matrix implies that for any kind of the disordered arrangement chains, it is equivalent to a lattice chain which contains a completely periodic array of tetratomic cells followed by the periodic array of diamond cells. On the other hand, the effective intra-cell hoppings of the two tetratomic and diamond building blocks are the same for any energy eigenvalues (Eqs. \ref{eqirs} and \ref{energy-dependent-term}). Thus, we could say this amorphous system with certain parameters equals to the periodic chains with four continuous spectrum domains (energy bands) and all the extended Bloch bulk eigenstates, which is counterintuitive. On the other hand, the Bloch bands are topological and topological in-gap states are localized at the edges. In short, one can see that the translational symmetry is a sufficient but not necessary condition for the topological band phases.

%In short, according to this diamond-tetratomic model, we demonstrate that the topological band insulators in the amorphous systems, where all the bulk states are extended and topological in-gap states are localized at the edges. Thus, the translational symmetry is a sufficient but not necessary condition for the topological band states. 

\begin{figure}[hbtp]
\centering 
\includegraphics[width=0.45\textwidth]{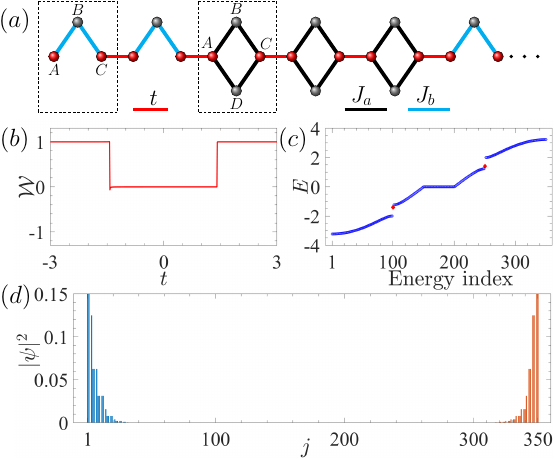}
\caption{The diamond-trimer model and its topological features. $(a)$ A typical realization of the diamond-trimer model. $(b)$ Numerically calculate winding number $w$ as a function of intre-cell hopping $t$, with the intra-cell hopping $J_a=1$, $J_b=\sqrt{2}$. $(c)$ The real-space energy spectrum of the original diamond-trimer model with the system parameters $J_a = 1$, $J_b=\sqrt{2}$, $t = 2$, and the number of building cells $N=100$. $(d)$ The probability density of the topological edge states with energy $E = -\sqrt2$, marked as red diamond in $(c)$, localized on the boundary of the system.} 
\label{FigTrimer}
\end{figure}

\emph{The diamond-trimer model}---
Motivated by the above observations, we now provide another example about the topological band insulators in the amorphous systems. As shown in Fig.\ref{FigTrimer}(a), the building blocks are diamond and trimer cells, which are arranged with the equal probability in a random fashion. The intra-cell hopping amplitudes of the diamond (trimer) lattice are uniformly set to $J_{a}$ ($J_{b}$), and $t$ is the inter-cell coupling amplitude. The ratio of intra-cell hoppings $J_a$ and $J_b$ is chosen as $J_b=\sqrt{2}J_a$.

We choose lattice sites $A$ and $C$ as the structural set $S$, and the other sites as the complement set  $\overline{S}$. The effective 1D dimerized model reduced by ISR have the  energy-dependent intra-cell hopping amplitudes $J(E)=\frac{2J_a^{2}}{E}$ and on-site potentials $V(E)=V_{A}(E)=V_{C}(E)=\frac{2J_a^{2}}{E}$.  Under PBC, we can get that the isospectral effective Hamiltonian $\widetilde{H}_S(E, k)$ in momentum space
\begin{eqnarray}
\widetilde{H}_S(E, k)=\left[
  \begin{matrix}
    V(E) &  J(E)+te^{-ik}   \\
    J(E)+te^{ik} & V(E)  \\
  \end{matrix}
  \right].
  \label{matrix4}
\end{eqnarray}

For the Hamiltonian of this effective dimerized model, $\widetilde{H}_S(E, k) - V(E)I$ takes on the mathematical form of the latent Su-Schrieffer-Heeger (SSH) model with chiral symmetry\cite{RontgenM24PRB}. Thus, the topological properties and phase transitions of the diamond-trimer lattice model could be analyzed by the winding number $\mathcal{W}$ (Eq.\ref{eq2.3}). In Fig.\ref{FigTrimer}$(b)$, we plot the winding number $\mathcal{W}$ for the lowest band (at $2/7$ filling) of the system as a function of inter-cell hopping $t$, with the parameters $J_{a}=1$ ($J_{b}=\sqrt{2}J_{a}$). For the $|t| > \sqrt 2|J_a|$ case, the amorphous system is in topological band insulators phase. We further numerically calculate the energy spectrum and eigenstates with Hamiltonian parameters $J_a=1$, $J_b=\sqrt 2$, and $t=2$, as shown in Fig.\ref{FigTrimer}$(c)$. The topological in-gap states with the energy $E=\pm\sqrt2$ is explicitly highlighted in the spectrum (marked by a red diamond symbol) and its spatial localization on the edge of system is further validated through the calculated probability density (see Fig.\ref{FigTrimer}[d]).

For the diamond and trimer cells, the intra-cell hoppings $J_b=\sqrt{2}J_a$, it is easy to show that the string transfer matrices $M_{dia}=M_{tri}$ ($[M_{dia},M_{tri}]=0$). Thus, this amorphous system reduces to a periodic chain with energy dispersion $E^3-(t^2+4J_a^2)E-4tJ_a^2\cos{k}=0$ and contains three Bloch bands. It should be mentioned that there exists a flat band at $E=0$, whose eigenstates are compact localized states (CLSs) emerged from the symmetric diamond cells. These macroscopically degenerate CLSs bound states lie in the continuum of extended states of the second (center) band, forming bound states in the continuum (BICs)\cite{HsuCW16NRM}. According to the local (permutation) symmetry, the diamond cell could be divided into a trimer with intra-cell hopping $\sqrt2 J_a$ and isolated site with potential $E=0$\cite{RontgenM18PRB}. As a result, the disordered whole system can be viewed as a hybrid of a disordered arrangement chain with two different intra-cell hoppings ($\sqrt2 J_a$ and $ J_b$) trimer cells and a set of $N/2$ isolated sites. With the intra-cell hoppings $J_b=\sqrt2J_a$, the disordered arrangement chain becomes periodic lattice processing three Bloch bands, despite the lack of translational symmetry.

\emph{Conclusion}---
In summary, we demonstrate that the translational symmetry is not always a necessary condition for the topological band states. By way of the ISR approach, we take the two amorphous systems as examples and propose a flexible construction method of the topological band insulators, where the topological in-gap edge states and all extended bulk Bloch states appear. On the other hand, we illustrate the fascinating phenomenon of the flat band and macroscopically degenerate bound states in the continuum. Additionally, these unexpected phenomena in our work can be identified in artificial systems such as photonic systems, phononic systems, and topo-electrical circuits. The construction of the topological band metals and superconductors in the amorphous systems is an important future direction. In one word, our findings uncover an arena for the construction and exploration of the topological band states beyond the translational symmetry systems paradigm.

%Although the Tetris pieces have different structures/shapes, they can perfectly fill the two-dimensional space.

\emph{Acknowledgements.}---We thank the useful discussion with Malte Röntgen, Feng Liu, and Wenlong Gao. This work was supported by the National Natural Science Foundation of China (Grant No. 12074101) and the Natural Science Foundation of Henan (Grant No. 212300410040).

%\bibliography{TopologicalInsulators}
%apsrev4-2.bst 2019-01-14 (MD) hand-edited version of apsrev4-1.bst
%Control: key (0)
%Control: author (8) initials jnrlst
%Control: editor formatted (1) identically to author
%Control: production of article title (0) allowed
%Control: page (0) single
%Control: year (1) truncated
%Control: production of eprint (0) enabled
%

 \end{document}